\documentclass[twocolumn]{aastex62}

\usepackage{amsmath,amssymb,amstext}
\usepackage{subfigure}

\usepackage{float}

\usepackage[utf8]{inputenc}
\usepackage{lmodern}

\usepackage{mathtools,nccmath}%

\usepackage{environ}

\usepackage{natbib}
\usepackage{hyperref}

\usepackage{xcolor}

\newcommand{\muf}{\mu_{\rm f}}

\newcommand{\rj}{R_{\rm jet}}

\newcommand{\geff}{\Gamma_{\rm eff}}
\newcommand{\Emax}{\mathcal{E}_{\rm max}}
\newcommand{\maxg}{\rm max(\Gamma_{\rm pr})}

\begin{document}

\title{\emph{Espresso} and Stochastic Acceleration of Ultra-high-energy Cosmic Rays in Relativistic Jets}

\author[0000-0001-9475-5292]{Rostom Mbarek}
\email{rmbarek@uchicago.edu}
\affiliation{University of Chicago, Department of Astronomy \& Astrophysics, 5640 S Ellis Ave., Chicago, IL 60637, USA}
\affiliation{Kavli Institute for Cosmological Physics, The University of Chicago, Chicago, IL 60637, USA}
\affiliation{Enrico Fermi Institute, The University of Chicago, Chicago, IL 60637, USA}

\author[0000-0003-0939-8775]{Damiano Caprioli}
\affiliation{University of Chicago, Department of Astronomy \& Astrophysics, 5640 S Ellis Ave., Chicago, IL 60637, USA}
\email{caprioli@uchicago.edu}
\affiliation{Enrico Fermi Institute, The University of Chicago, Chicago, IL 60637, USA}

\begin{abstract}

In the \emph{espresso} scenario, ultra-high-energy (UHE) cosmic rays (CRs) are produced via a one-shot reacceleration of galactic-like CRs in the relativistic jets of active galactic nuclei, independently of the scattering rate dictated by  magnetic fluctuations.
In \cite{mbarek+19}, we traced test-particle CRs in high-resolution magnetohyrodynamic (MHD) jet simulations and found that the associated spectral slope, chemical composition, and anisotropy are consistent with  UHECR phenomenology. 
In this work, we extend such an analysis by including sub-grid pitch-angle scattering to model small-scale magnetic turbulence that cannot be resolved by MHD simulations. 
We find that a large scattering rate unlocks stochastic acceleration and fosters the energization of lower-energy CRs, which eventually leads to harder UHECR spectra. 
Yet, the particles that achieve the highest energies (up to the Hillas limit) are invariably produced by \emph{espresso} acceleration and their spectrum is independent of the assumed sub-grid scattering rate.
\end{abstract}

\section{Introduction\label{intro}}
The mechanism responsible for the acceleration of ultra-high energy cosmic rays (UHECRs, above $10^{18}$eV) is still much debated. 
Acceleration models often hinge on simple back-of-the-envelope estimates of the maximum energy  achievable in a particular environment, the so-called \emph{Hillas criterion} \citep[][]{cavallo78,hillas84}. 
Astrophysical sources such as $\gamma$-ray bursts \citep[e.g.,][]{vietri95,waxman95}, tidal disruption events \citep[e.g.,][]{farrar+14}, newly-born millisecond pulsars \citep[e.g.,][]{blasi+07,fang+12}, and active galactic nuclei \citep[AGNs; e.g.,][]{ostrowski00,osullivan+09,murase+12,matthews+19} have all been suggested as potential acceleration sites \citep[see, e.g.,][for a review]{kotera+11}. 
However, the Hillas criterion is a \emph{necessary} but not \emph{sufficient} condition \cite[also see][for a condition on the source luminosity]{Waxman04}.
A proper theory of UHECR acceleration should account for: 
i) when and how particles are injected into the acceleration site;
ii) the mechanism through which acceleration occurs;
and iii) the properties of the released particles (spectrum, chemical composition, anisotropy, ...).
Also, such a theory should be general, in the sense that results cannot be fine-tuned to model/environmental parameters, and be comparable with observations. 
Such a theory is still missing. 

In a series of papers, we are investigating a promising theoretical framework that has the potential to satisfy the requirements above.
\citet{caprioli15} outlined the so-called \emph{espresso} mechanism, a model-independent form of acceleration for UHECRs in AGN jets, that was later confirmed both analytically \citep{caprioli18} and numerically \citep[][hereafter, MC19]{mbarek+19}. 
The basic idea is that CR \emph{seeds} accelerated in supernova remnants should be able to penetrate into relativistic jets, tap the motional electric field in the ultrarelativistic jet spine, and generally receive a boost of a factor of $\sim \Gamma^2$ in energy, where $\Gamma$ is the Lorentz factor of the relativistic flow. 
In a nutshell, \emph{espresso} acceleration is analogous to inverse-Compton scattering, with CR seeds and the jet magnetic field substituting for photons and relativistic electrons. 
In the \emph{espresso} framework, one (sometimes two) acceleration cycles saturate the particle energy gain, different from typical stochastic processes where each cycle grants a fractional energy gain, and the number of cycles is controlled by a prescribed parameter, such as a scattering rate.
Unlike in stochastic acceleration mechanisms, in the \emph{espresso} case: i) there are no multiple cycles involved where acceleration arises from a balance between energy gains and losses, and ii) there are no prescribed microphysical (often unconstrained) parameters, such as the diffusion rate.

With sufficiently large Lorentz factors  \citep[$\Gamma\gtrsim 30$, as inferred from multiwavelength observations of powerful blazars, e.g.,][]{tavecchio+10,zhang+14}, one \emph{espresso} shot would be sufficient to transform the highest-energy galactic CRs at $10^{17}$~eV to the highest-energy UHECRs at $\sim 10^{20}$~eV.
Such a re-acceleration process also naturally reproduces the measured UHECR chemical composition \citep{auger14a,auger17b,dembinski+18,heinze+19}, which is proton-like at $10^{18}$~eV and increasingly heavier at higher energies.
Also see \cite{pierog13,abbasi+15} for a comparison between data from Auger and Telescope Array.

\subsection{\emph{Espresso} Acceleration in MHD Jets}

In MC19 we explored the \emph{espresso} framework by following a bottom-up approach that keeps parametrizations at a minimum. More specifically, we studied how particles are injected and accelerated by propagating test seeds in state-of-the-art 3D magnetohyrodynamic (MHD) simulations of relativistic jets. We found that particles gain energy to reach the Hillas limit of the jet through one/two \emph{espresso} shots \emph{without} any added scattering. In this regard, \emph{espresso} is model-independent and does not rely on assumptions on the jet structure and the diffusion rate. In general, we found no evidence that particles at the highest energies undergo any type of stochastic acceleration. We then concluded that \emph{espresso} acceleration is indeed generic and that particles can experience more than one \emph{espresso} shot even in low-$\Gamma$ jets.

Our high-resolution MHD simulations \citep[run with PLUTO, see ][]{mignone+07,rossi+08}, which use adaptive mesh refinement \citep{mignone+12}, can capture large-scale magnetic fluctuations self-consistently.
Still, they cannot account for the potential role of smaller-scale turbulence, below the grid resolution, in particle scattering.
In this paper we introduce sub-grid scattering (SGS) in the whole computational domain via a Monte Carlo approach: we propagate particles on the grid with a standard Boris pusher \cite[e.g.,][]{birdsall+91}, while prescribing a finite probability per unit time for particles to change their pitch angle with respect to the local magnetic field. 
Introducing SGS effectively includes an additional stochastic process with respect to our previous analysis, so we refer to the extra acceleration that one may obtain as (a type of) stochastic acceleration. 
Stretching the inverse-Compton analogy even further, adding SGS makes the environment more Compton-thick and fosters the comptonization of the seeds, without changing the maximum achievable energy.

This approach allows us to investigate the role of stochastic acceleration, which is fostered by increased scattering, in realistic AGN jets.
Understanding the properties of the resulting spectra that depend on the assumed level of SGS is particularly important because the actual diffusion rate in different jet regions is effectively a free parameter in any model and can hardly be constrained by observations.
In particular, we address the following questions:
\begin{itemize}
\item Does SGS have an impact on the fraction of seeds that can be reaccelerated?
\item Do particles typically gain more energy in the presence of SGS? 
\item Does the maximum achievable energy depend on the scattering rate? 
\item What is the relative importance of \emph{espresso} and stochastic acceleration?
\end{itemize}
We find that adding SGS increases the percolation rate into the high-$\Gamma$ regions of the jet and helps to break the correlation between in- and out-going angles \citep{caprioli15,caprioli18}, and hence facilitates multiple \emph{espresso} acceleration cycles.
Both effects go in the direction of increasing the acceleration efficiency; 
therefore, the results of MC19 can be seen as a \emph{lower limit} on the effectiveness of the jet at accelerating particles. 

Furthermore, adding SGS fosters stochastic shear acceleration at the jet/cocoon interface  as well as diffusive shock acceleration in the jet backflows, two energization mechanisms potentially responsible for the production of UHECRs, as suggested, e.g., by \cite{ostrowski98,ostrowski00,liu2+17,fang+18,kimura+18,tavecchio21} and \cite{osullivan+09,matthews+19}, respectively.

The plan of the paper is the following: 
in \S\ref{init_traj} we analyze  the effects of SGS on trajectory patterns, energy gains, and energy spectra. 
In \S\ref{comparison}, we discriminate between particles that underwent stochastic or \emph{espresso} acceleration and assess the maximum energy that can be achieved through each mechanism, showing how the highest-energy particles are invariably \emph{espresso} accelerated. 
Finally, we summarize the implications of our results for astrophysical applications in \S\ref{realistic}.

\section{trajectories and spectra of released particles}\label{init_traj}

We carry out our analysis using 3D relativistic MHD simulations performed with PLUTO, which includes adaptive mesh refinement \citep{mignone+12,rossi+08}.
We consider the same initial conditions as in MC19 in order to single out the role of small-scale scattering on top of a known situation (Please refer to MC19 for more details on the MHD simulation setup and properties of the jet).
The jet is launched with Lorentz factor $\Gamma_0 = 7$ along $\hat{z}$ through a nozzle with a magnetization radius $\rj$ in a box that measures 48$\rj$ in the $x$- and $y$-directions and 100$\rj$ in the $z$-direction in a grid that has $512 \times 512 \times 1024$ cells with four refinement levels. 
The initial conditions are set by the jet/ambient density contrast $\psi$, the jet sonic and Alfv\'enic Mach numbers $M_s\equiv c/c_s$, and $M_A\equiv c/v_A$, where $c$, $c_s$, and $v_A$ are the light, sound, and Alfv\'en speed, respectively. 
Our initial parameters are $\psi = 10^{-3}$, $M_s = $3, and $M_A = 1.67$.

\begin{figure}
\centering
\includegraphics[width=0.48\textwidth,clip=true,trim= 0 0 0 0]{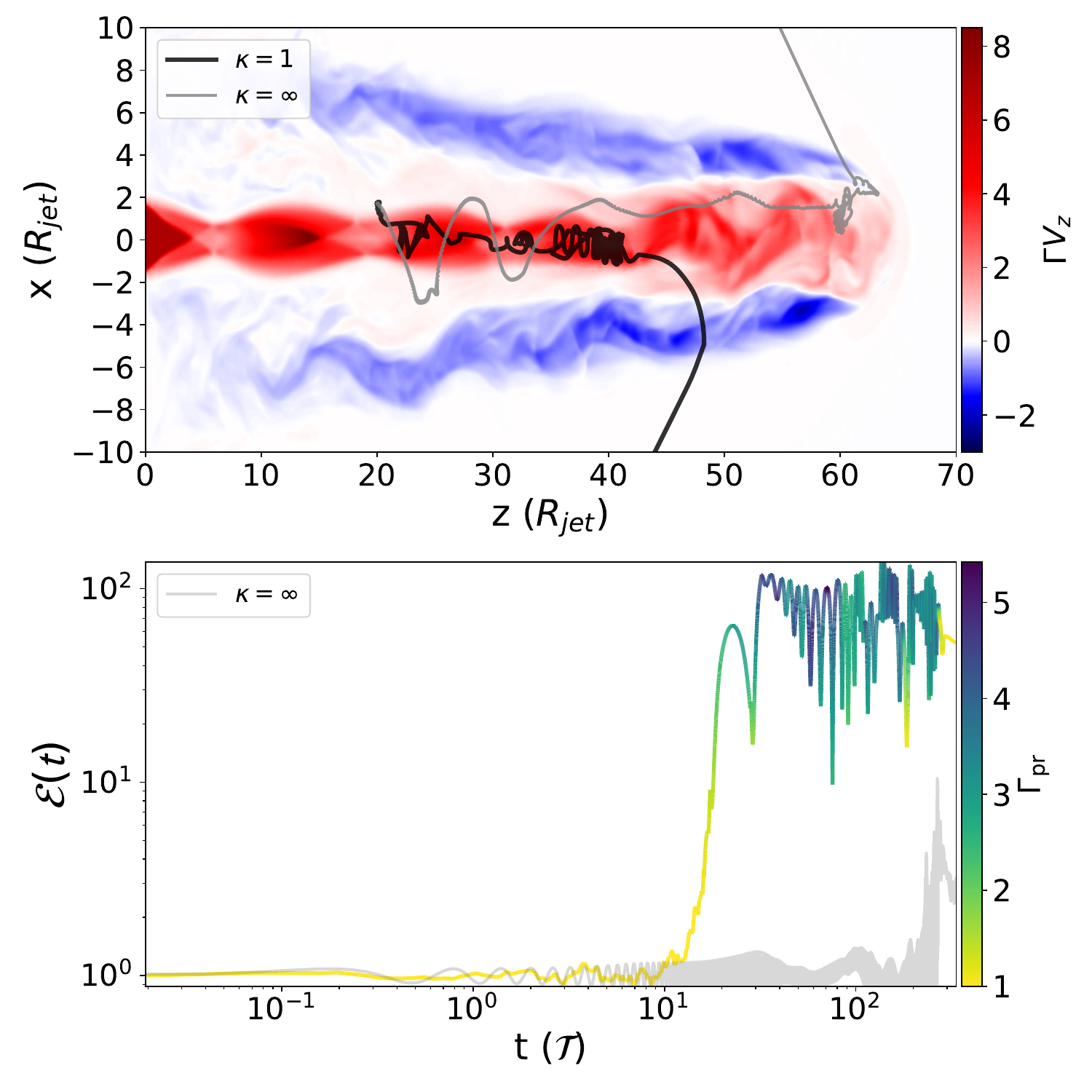}
\caption{Trajectory and energy gain for representative particles with $\kappa =1$ and $\kappa =\infty$.
Top panel: particle trajectories with the same initial conditions overplotted on the 4-velocity component $\Gamma v_z$ of the flow.
Bottom panel: energy evolution, color coded with the instantaneous Lorentz factor probed, $\Gamma_{\rm pr}$ for the $\kappa =1$ particle as a function of the of the gyroperiod $\mathcal T \equiv \frac{2 \pi m \gamma_i}{q B_0}$ as defined in the text. 
Such a particle initially gains a factor of $< 4$ in energy through stochastic acceleration and then experiences multiple \emph{espresso} shots in the high-$\Gamma$ jet regions. 
The particle with $\kappa =\infty$ (grey line), on the other hand, only experiences one \emph{espresso} shot.  Both particles are initialized with $\alpha_{\rm i} \sim 0.075$. }
\label{traj}
\end{figure}

As in MC19, we consider two possible orientations of the jet toroidal magnetic field, $B_\phi\lessgtr 0$, which physically corresponds to a current $J_z\lessgtr 0$ along the jet axis (case A and B, respectively).
These cases lead to similar UHECR spectra, and differ only for the angular distribution of the released particles (Figure 11 in MC19).
When not specified, we plot results for case A only; 
a synthesis on the differences and similarities between both cases is included in \S\ref{anisotropy}.

We propagate $\sim$100,000 test-particles in this jet with a wide range of initial gyroradii $\mathcal R$ and positions, with the same initial conditions as in MC19 and for the scattering rates discussed in \S\ref{bohm_diffusion}. Particles are homogeneously and isotropically initialized from linearly spaced locations $(r_{\rm i}, \phi_{\rm i}, z_{\rm i})$ around the spine of the jet where $r_{\rm i}/\rj\in [0.2,5]$, $z_{\rm i}/\rj\in [2,60]$ and $\phi_{\rm i}\in [0,2\pi]$. Additionally, their initial pitch angles are also linearly spaced to span all possible angles.
It is useful to introduce the particle gyroradius normalized to the jet radius in the reference value of the magnetic field, $B_0$ as
\begin{equation}
    \alpha(E,q)\equiv \frac{\mathcal R(E,q)}{\rj},
\end{equation}
where $E$ and q are the particle's energy and charge, respectively. The particles' injection spectrum is flat in the interval of initial gyroradii $\alpha_{\rm i}\in [2 \times 10^{-4}, 10^{1}]$, where $\alpha_{\rm i}$ goes beyond the jet's Hillas limit, thus probing an extended energy range. Since trajectories only depend on the rigidity $E/q$,  our particles are effectively representative of different nuclei.
Note that, since the rigidity scales with the jet radius, there is no absolute energy scale for the particles. 
When typical values for $B_0$ and $\rj$ are chosen, the energies of the considered particles are close to those of UHECRs: 
for instance, $B_0=100\mu$G and $\rj=100$pc would push the Hillas limit to $\sim 10^{20}$eV for iron nuclei \citep{caprioli15}.

For the given resolution of about 10 cells per $\rj$, our MHD simulations resolve magnetic fluctuations down to scales comparable to the gyroradius of particles with $\alpha\gtrsim 0.1$.
Note that particles can be propagated even if their gyroradius is smaller than the grid size, as long as a rigidity-dependent timestep sufficient to resolve their gyration is used.
The role of SGS is to model the effect of unresolved magnetic structures for such particles.

\subsection{Pitch-angle scattering due to unresolved turbulence \label{bohm_diffusion}}
The simulations in MC19 do not include any SGS and hence can be seen as a limiting case with minimum scattering. 
The opposite limit would be to assume that diffusion occurs in the Bohm regime, in which the mean free path for pitch-angle scattering is as small as the particle's gyroradius. 
We span across these two regimes by introducing a diffusion coefficient $D$ that is a function of the particle rigidity and the local magnetic field that reads:
\begin{equation}\label{eq:bohm}
    D(\mathcal{R})\equiv \frac{\kappa}{3} c \mathcal{R}(E, q,B),
\end{equation}
where $\mathcal{R}$ is the particle gyroradius and $\kappa$ defines the number of gyroradii per sub-grid scattering \citep[sometimes called the gyrofactor, e.g.,][]{amano+11}. 
$\mathcal{R}$ is a function of the local magnetic field $B(\textbf{r})$, which in simulations is scaled to the initial value at the magnetization radius $B(r=\rj)\equiv B_0$ (see MC19 for details).

In this work, we span a wide range of SGS rates by posing  $\kappa = 1000,$ 100, 10, and 1 (Bohm diffusion) and investigate their effects on particle acceleration, energy gains, and anisotropy;
we compare these cases with the results from MC19, which do not include SGS and correspond to $\kappa = \infty$. 
For Alfv\'enic fluctuations, $\kappa$ is related to the power in modes with wavenumber resonant with the particle gyroradius ($k\mathcal R\sim 1$);
Bohm diffusion corresponds to the case in which $\delta B(k)\sim B_0$, while for $\delta B(k)<B_0$ one has that $\kappa\propto [B_0/\delta B(k)]^{2}$ \citep[e.g.,][]{skilling75a}.
In general, the exact relation between diffusion coefficient and magnetic power spectrum depends on the nature of the unresolved turbulence \citep[spectrum, anisotropy, helicity, see e.g.,][]{schlickeiser02}.

Pitch-angle diffusion may be accompanied by diffusion in momentum space due to the finite velocity of the scattering centers  \citep[e.g.,][]{skilling75a,ptuskin88}.
We do not include this kind of second-order Fermi acceleration here because it is expected to be underdominant in super-Alfv\'enic flows and because its efficiency is reduced for particles with gyroradii larger than the largest waves in the system, for which  diffusion becomes almost independent of energy \citep{osullivan+09}.

\subsection{Particle trajectories with scattering}

Intuitively, adding SGS is expected to facilitate the diffusion of seed particles in and out of the jet spine, the region with ultra-relativistic flows where \emph{espresso} acceleration occurs.
To illustrate that this is actually recovered in simulations, in Figure~\ref{traj} we show a representative trajectory of a particle with $\kappa = 1$, i.e., experiencing Bohm diffusion;
the bottom panel shows the particle energy gain $\mathcal{E}$ as a function of its relativistic gyroperiod $\mathcal T \equiv \frac{2 \pi m \gamma_i}{q B_0}$ where $\gamma_i$ is the initial Lorentz factor, $q$ the charge, and $m$ the mass.

To assess the role of SGS, we overplot the trajectory of a particle with the same initial conditions but with $\kappa=\infty$ and $\alpha_{\rm i} \sim 0.075$. 
The $\kappa = 1$ particle in Figure~\ref{traj} follows a more jagged trajectory before entering the spine (where $\Gamma \gtrsim 2$) and experiencing \emph{espresso} gyrations. 
The reference particle, with the same initial conditions but $\kappa = \infty$, probes a maximum $\Gamma \sim 2$ and leaves the jet with a final energy gain of $\mathcal E \sim 3$ (energy diagram in grey in the bottom panel). 

\begin{figure}
\centering
\includegraphics[width=0.47\textwidth,clip=false,trim= 0 0 0 0]{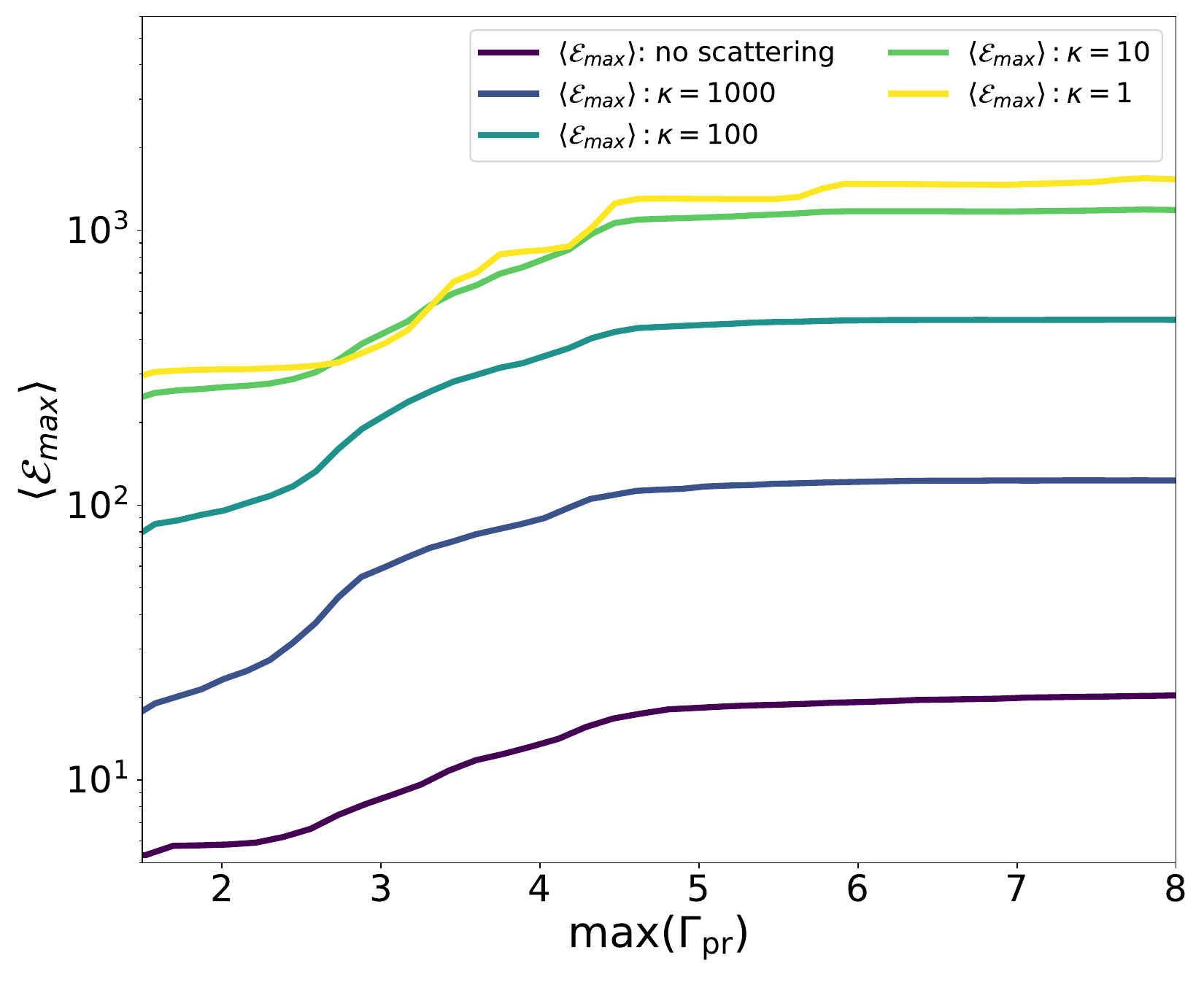}
\caption{Maximum energy gain $\langle \Emax  \rangle$ averaged on particles with $\mathcal{E} \geq 2 $ as a function of the maximum  Lorentz factor probed in the flow, $\Gamma_{\rm pr}$. Compare with Figure~6 in \citet{mbarek+19} for more details.}
\label{max_gain}
\end{figure}

\begin{figure}
\centering
\includegraphics[width=0.48\textwidth,clip=false, trim= 50 0 0 0]{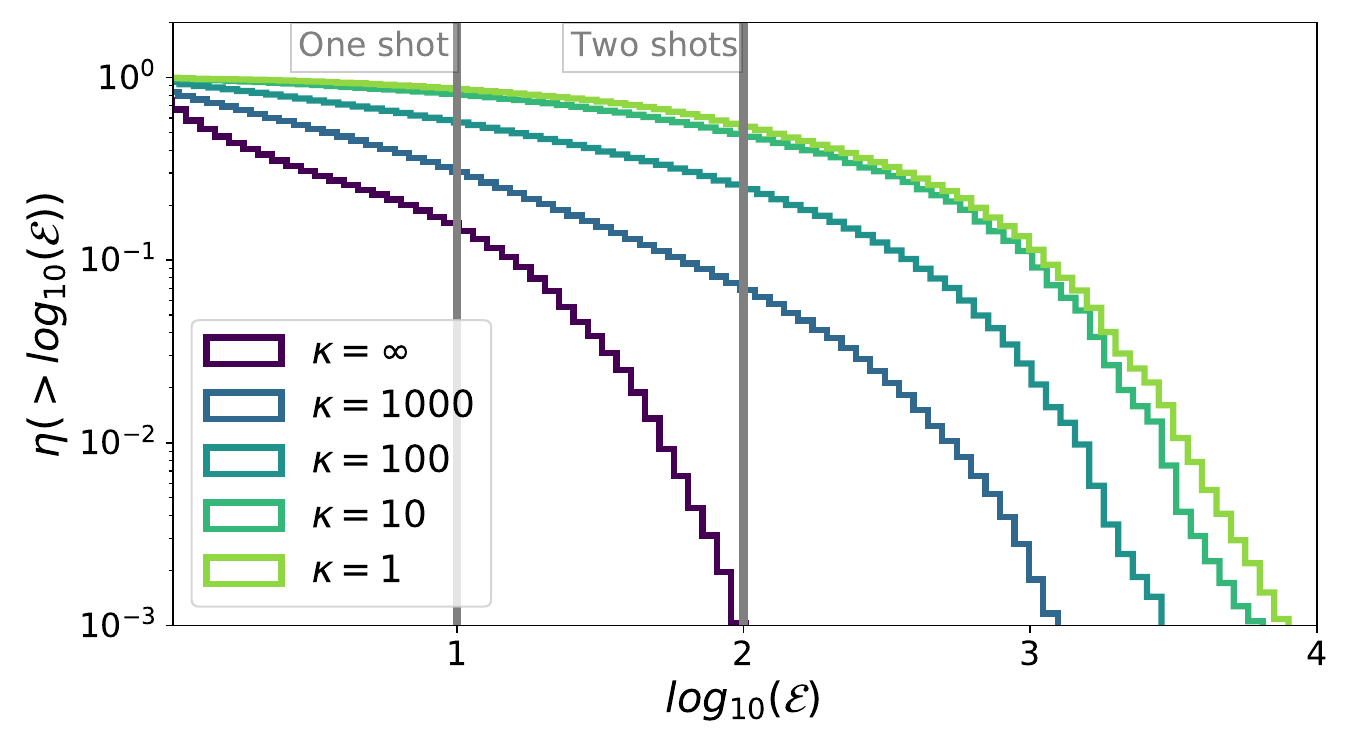}
\caption{Cumulative distribution of the energy gains of particles with a flat injection spectrum in $\eta$ such that $\alpha_{\rm i}\in [2 \times 10^{-4},2 \times 10^{-1}]$ for different SGS rates; $\eta$ is defined such that $\eta = \frac{\mathcal{E}}{N}  \frac{dN}{d \mathcal{E}}$ where N is the number of particles. The two vertical lines correspond to single- and double-\emph{espresso} shots in the effective jet Lorentz factor. Note how increasing SGS allows a larger fraction of seeds to be accelerated to higher energies. }
\label{ratio}
\end{figure}

We notice that the two particles have a similar evolution until  $t \sim 10 \mathcal{T}$, with energy oscillating within a factor of $\leq 2$.
Then, the $\kappa=1$ particle gets some further acceleration while still in the low-$\Gamma$ region (yellow color code in the bottom panel of Figure \ref{traj}) and penetrates the high-$\Gamma$ flow,  gaining a factor $\mathcal{E} \sim 120$ in energy by virtue of one canonical \emph{espresso} shot, i.e., the energy gain occurs over a single gyration.
This behavior is common to many particles and suggests that SGS leads to a type of stochastic acceleration at the jet-cocoon interface, which fosters particles to access the high-$\Gamma$ region and eventually get boosted via the \emph{espresso} mechanism. 

Finally, we point out that even in the case of Bohm diffusion, particles that enter the spine experience gyrations reminiscent of those observed without scattering, which are themselves similar to the analytical ones described in MC19 and \cite{caprioli18}.

In the next sections we quantify how SGS mostly affects: 1) the fraction of particles that can be \emph{espresso} accelerated and 2) the number of \emph{espresso} cycles that a particle may undergo.

\subsection{Energy gain and injection efficiency}

We have previously found that the maximum  energy gain of particles $\mathcal{E}_{\rm max}$ correlates with the maximum Lorentz factor that they probe along their trajectory  (Figure 6 in MC19). 
This correlation is recovered also when SGS is introduced, as shown in Figure \ref{max_gain}, but the overall normalization of the average energy gain depends on $\kappa$. 
Increasing SGS leads to larger energy gains, which saturate for $\kappa\lesssim 10$. 
The difference with respect to $\kappa = \infty$ lies in the fact that SGS helps to break the correlation between in- and out-going angles, which yields a larger energy gain per cycle.

This claim is further reinforced by the cumulative distribution $\eta(> \log_{10}(\mathcal{E}))$ of the energy gains of particles with $\mathcal{E} \geq 2 $ shown in Figure~\ref{ratio} for particles with initial gyroradii $\alpha_{\rm i}\in [2 \times 10^{-4},2 \times 10^{-1}]$, where $\eta$ is defined as $\eta =\frac{\mathcal{E}}{N} \frac{dN}{d \mathcal E}$ . 
We note a clear increase in the energy boosts that particles experience with increasing SGS.
While for $\kappa = \infty$ having more than one \emph{espresso} shot is relatively rare, already for $\kappa = 1000$, about 7\% of particles gain a factor of $\geq \geff^4$ in energy, the equivalent of two \emph{espresso} shots. 
This fraction increases to about 24\%, 48\%, and 54\% for $\kappa = 100$, $\kappa = 10$, and $\kappa = 1$ respectively.
Given the uncertainties in the seed abundance and spectra in AGN hosts, such efficiencies should not be translated directly into a UHECR luminosity for given AGNs;
still, our exercise suggests that the more SGS is added, the easier it is to accelerate pre-existing energetic particles.
Note that the observed UHECR flux may be accounted for by reaccelerating as little as $\sim 10^{-4}-10^{-2}$ of the highest-energy galactic seeds \citep{caprioli15}, which happens even for large values of $\kappa$.

Finally, we note that particle injection does not increase arbitrarily with the SGS level, but rather saturates for $\kappa\lesssim 10$;
therefore, even fluctuations at the level of $\delta B/B_0\lesssim 0.3$ (see \S\ref{bohm_diffusion}) at resonant scales may yield a maximum efficiency in seed re-acceleration.

\begin{figure}
\centering
\includegraphics[width=0.49\textwidth,clip=true,trim= 10 0 0 0]{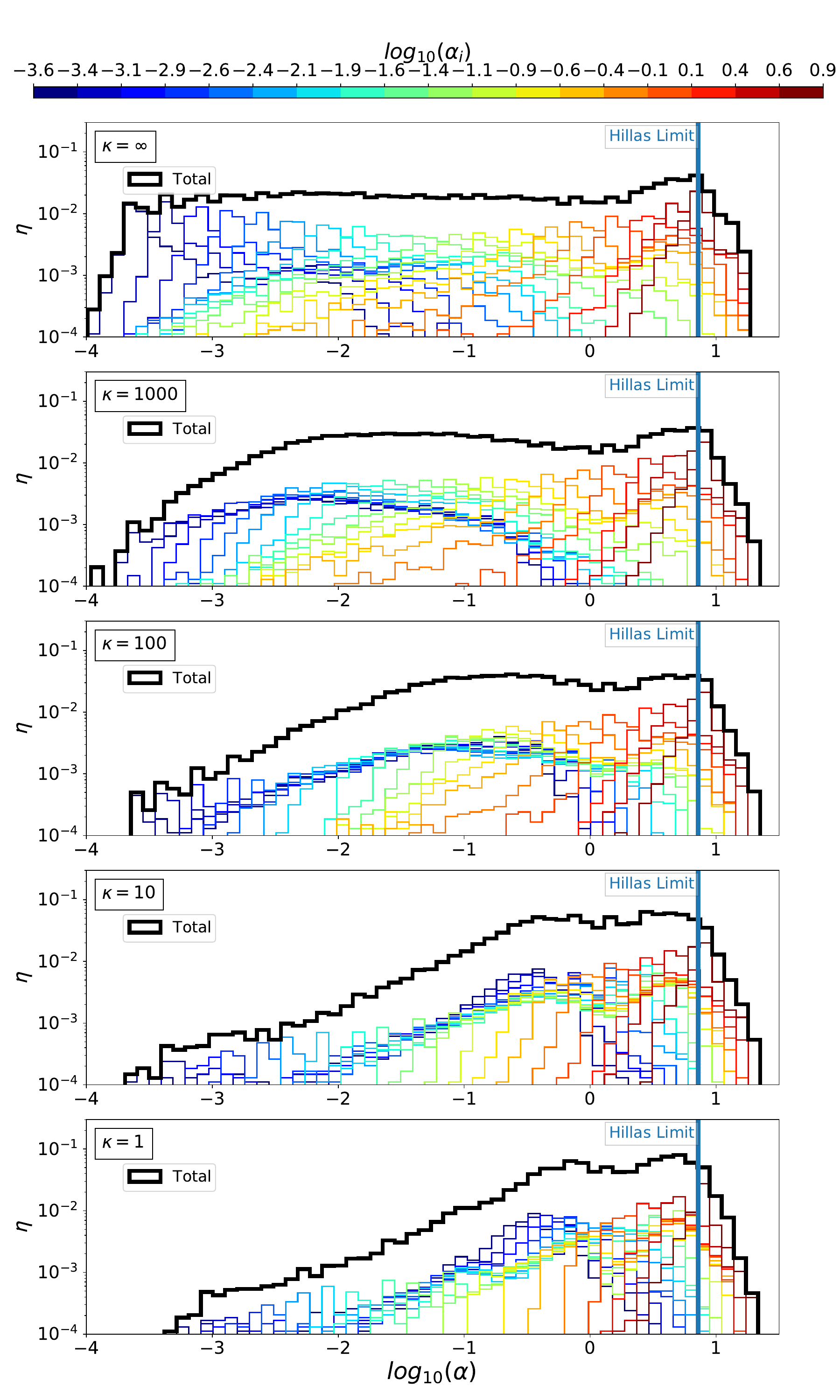}
\caption{ Distribution of the gyroradii of reaccelerated particles obtained for an injection spectrum flat in $\eta$ in the interval $\alpha_{\rm i} \in [10^{-3.6}, 8]$; $\eta$ is defined such that $\eta \propto \alpha \frac{dN}{d \alpha}$.
The thick black line shows the cumulative spectrum, while colored histograms correspond to initial gyroradii as in the color bar.
Seeds with $\alpha_{\rm i} \lesssim 1$ can undergo boosts as large as $\sim 50-100\gg\geff^2$, while for $\alpha_{\rm i} \gtrsim 1$ the energy gain is smaller and saturates at $\alpha_{\rm H}\approx 8$ (longitudinal Hillas criterion). }
\label{alpha_plot}
\end{figure}

\subsection{Effect of the scattering rate on the energy spectra}\label{sec_alpha}

Figure~\ref{alpha_plot} shows the distribution of the gyroradii of escaping particles,  $\alpha_{\rm f}$,  for different SGS prescriptions. 
The solid black line represents the total spectrum and is divided in the spectra produced by particles with a given initial $\alpha_{\rm i}$ (colored histograms). 
The blue vertical line marks the effective Hillas criterion for the jet considered: it corresponds to  $\alpha\sim 8$ because the average magnetic field in the spine region is $\sim 8 B_0$.

We have already discussed how particles in general undergo larger boosts with increasing SGS rates. 
Figure~\ref{alpha_plot} shows that in more detail by also capturing how the boost enhancement depends on the seed energy. 
The following features are worth noticing:

\paragraph{(i) Highest-energy particles.} 
For $ \alpha \gtrsim 2$ spectra are very similar regardless of the SGS rate, suggesting that the highest-energy particles are always \emph{espresso} accelerated. 
Close to the Hillas limit, stochastic acceleration, which depends on the level of SGS, does not contribute to either a larger fraction of particles or a larger maximum energy. 

\paragraph{(ii) Lower-energy particles.}
As SGS increases, the fraction of low-energy seeds that can be reprocessed by the jet grows considerably and energy gains of order of $\mathcal{E} \gtrsim 10^3$ (corresponding to two/three \emph{espresso} shots) are common;. 
The increase in the average energy gain shown in Figure~\ref{max_gain} and \ref{ratio} is driven by these particles.

\paragraph{(iii) Flattening of the spectrum.} 
For $\kappa=\infty$ the UHECR spectrum tends to be only slightly flatter than the injected one, showing a pile-up close to the Hillas limit. 
With SGS, the spectrum tends to become significantly flatter than the injected one, by about $\sim 0.9$ in slope for $\kappa = 1$.
Again, this effect is driven by a change in the behavior of the lower-energy particles rather than of the higest-energy ones.
This recovers results from Monte Carlo simulations of idealized jets, which report spectra as flat as $E^{-1}$ when Bohm diffusion is prescribed \citep{kimura+18}.
Note that rather hard injection spectra are favored by current models for the flux and composition of UHECRs \cite[e.g.,][]{gaisser+13,aloisio+14,taylor14}.

\subsection{Angular Distribution of Escaping Particles}\label{anisotropy}

Let us focus now on the anisotropy of the accelerated particles, which is important for assessing a possible correlation between the UHECR directions of arrival and local sources. 
In MC19, we found that without SGS particles preferentially move along the spine of the jet until $z \sim 60$ and get deflected in the cocoon after leaving the relativistic region. 
In that case, the sign of $B_\phi$ in the jet, which controls the sign of the motional electric field in the cocoon, either disperses ($E_r>0$, case A) or collimates ($E_r<0$, case B) the escaping particles.
In both cases the UHECR distribution is never beamed in an angle $1/\Gamma$ as one may expect from a relativistic source; 
in case A, where the radial (motional) electric field points outward, it is almost isotropic.

\begin{figure}
\centering
\includegraphics[width=0.48\textwidth,clip=false,trim= 0 0 0 0]{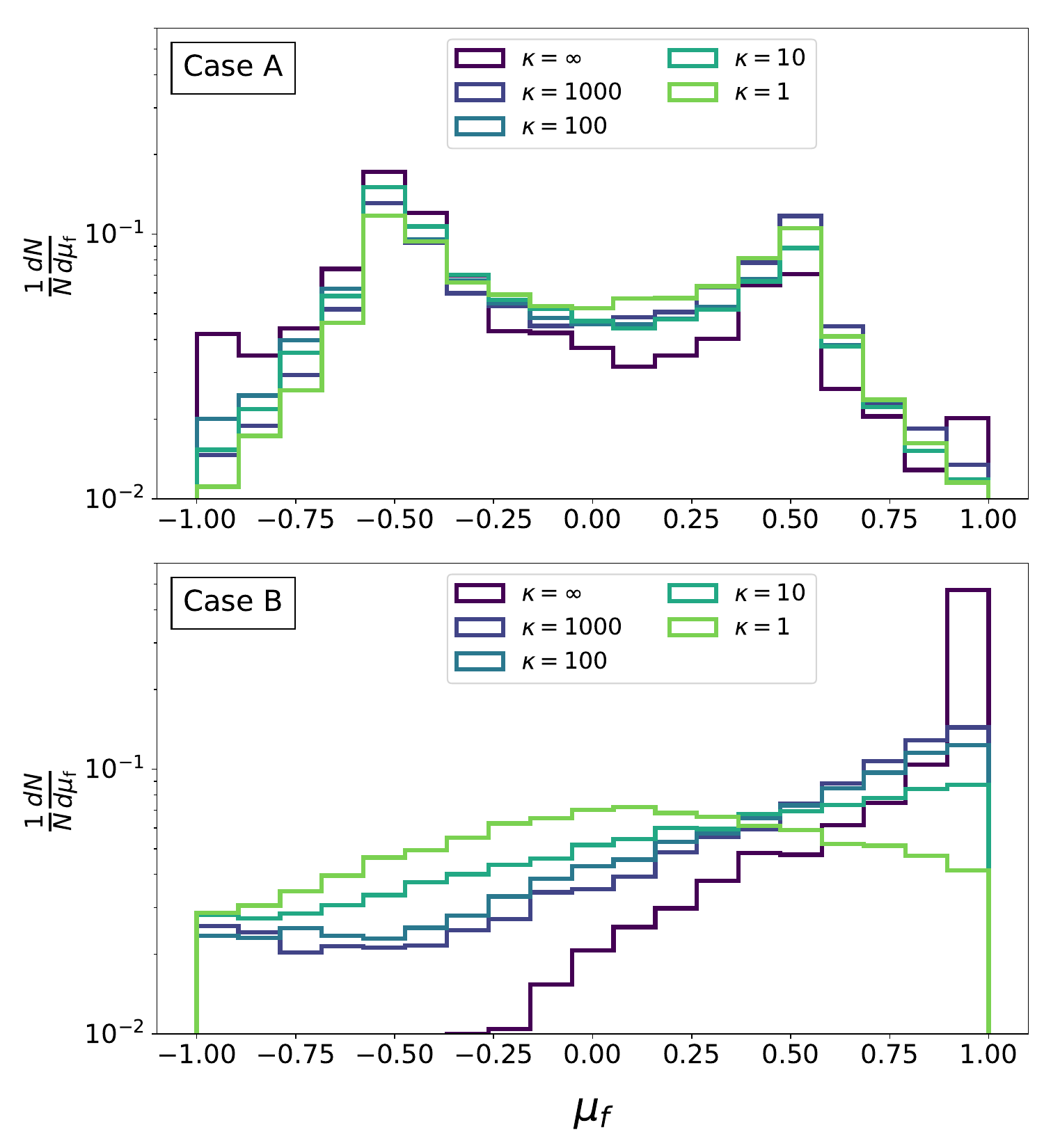}
\caption{Distribution of the cosines of the final angle of flight for accelerated particles with $\mathcal{E} \geq \geff^2$. 
In case A particles escape the jet isotropically independently of the scattering rate, while in case B they are less and less beamed as SGS rate increases.}
\label{isotropy}
\end{figure}

We repeat the same analysis here for all the particles that gain at least a factor of $\Gamma^2$ in energy in the presence of SGS. 
Figure~\ref{isotropy} shows the distribution of $\muf$, the cosine of the angle between the final particle velocity and the $z$-axis, for both case A and case B. 
While for case A the distribution is isotropic independently of $\kappa$, increasing the SGS leads to a progressively more isotropic angular distribution of escaping particles in case B, too (bottom panel of Figure~\ref{isotropy}).
This is a natural consequence of introducing additional pitch-angle scattering; it reinforces the idea that relativistic jets may be quite isotropic UHECR emitters and, as a result, we may expect comparable contributions from AGNs that we classify as blazars (with jets along the line of sight) or Fanaroff-Riley radio galaxies.

Since adding SGS induces an isotropization of the outgoing fluxes in both cases, we treat only case-A particles in the remainder of the paper for simplicity. 
Overall, we find no appreciable differences in particle trajectories or final spectra between the two cases.

\section{Espresso and stochastic acceleration}\label{comparison}

In \S\ref{init_traj} we have showed one instance in which adding SGS may foster injection into \emph{espresso} acceleration. 
Let us now discuss differences and interplay between \emph{espresso} and stochastic acceleration.

In general, stochastic acceleration relies on repeated crossings of the  shearing layers at the interface between the jet and the cocoon \citep[e.g.,][]{ostrowski98, ostrowski00,fang+18,kimura+18}, or repeated diffusive shock acceleration and/or turbulent acceleration in the cocoon \citep[e.g.,][]{osullivan+09,matthews+19};
\emph{espresso} acceleration, instead, relies on a few Compton-like scatterings against the most relativistic regions of the jet.

\subsection{Spectral signatures}\label{spec_sig}

\begin{figure}
\centering
\includegraphics[width=0.45\textwidth,clip=false,trim= 50 0 0 0]{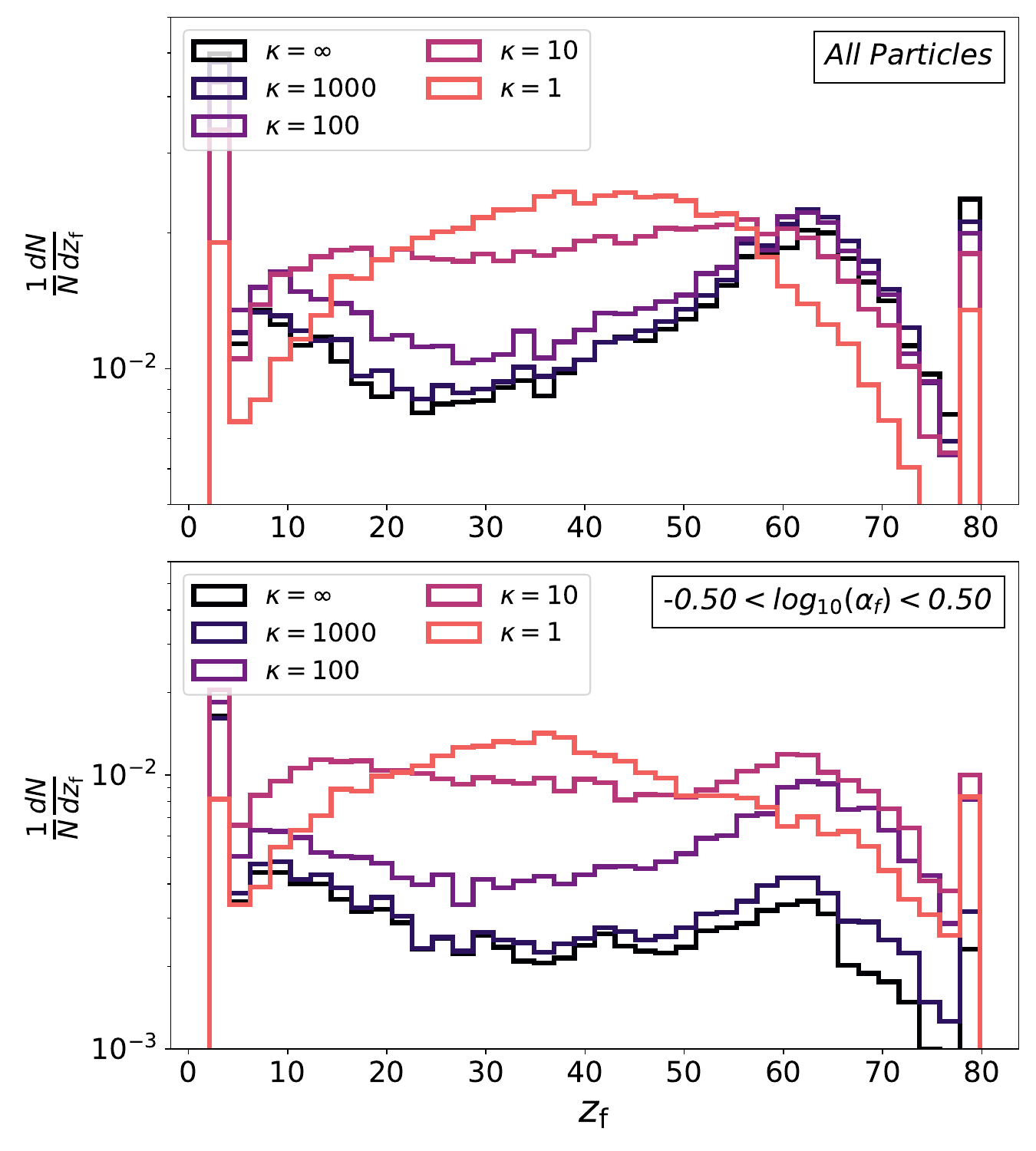}
\caption{Top Panel: Final $z$ position $z_{\rm f}$ of all particles after they escape the jet. 
Bottom Panel: as above, but only for particles that escape with energies around the stochastic Hillas limit. }
\label{zf}
\end{figure}

A natural question is whether the spectral diversity in the sub-Hillas region (see \S\ref{sec_alpha} and Figure~\ref{alpha_plot}) is due either to a more effective injection of seeds into \emph{espresso} acceleration or to stochastic acceleration.
As long as relatively-low $\Gamma$ factors are involved, the distinction between the two processes is more semantic than factual, in the sense that many acceleration events are invariably needed to achieve large energy gains.

For simplicity, we define the contribution of stochastic acceleration as the contribution to the UHECR spectrum that exceeds the one provided by the $\kappa=\infty$ case, i.e., the case with large-scale turbulence  in which particles gain energy only via a few interactions with the highly-relativistic jet spine.
Figure~\ref{alpha_plot} shows that adding SGS leads to a more and more prominent secondary peak below the Hillas limit that moves towards larger energies ($ \alpha \sim 7\times 10^{-3}, 7\times 10^{-2}, 0.6$) for $\kappa =$1000, 100, 10 respectively. 

We interpret such peaks as due to diffusive escape from the side of the jet, which depends on the level of SGS.
The diffusion length $\lambda$ of the particles in the radial direction (transverse to the jet axis) can be expressed as:
\begin{equation}\label{eq:diff}
    \lambda_{d} \approx \frac{D(\alpha)}{V_r},
\end{equation}
where $V_r$ is the radial velocity component of the jet. 
Averaging over $V_r$ in our fiducial simulation returns $\langle V_r \rangle \approx 0.01$ and the diffusion length becomes:
\begin{equation}\label{eq:diff_len}
    \lambda \sim 30 \frac{B_0}{B_{\rm coc}} \alpha \kappa ~\rj,
\end{equation}
where $B_{\rm coc} \sim 2 B_0$ is the averaged magnetic field in the cocoon. 
The maximum rigidity that a particle can achieve via stochastic acceleration is set by equating such a diffusion length with the typical transverse size of the shearing region, i.e., $\lambda(\mathcal{R_{\rm max}}) \sim 10 \rj$ (see Figure \ref{traj}).
This simple scaling reproduces the position of each sub-Hillas peak in Figure~\ref{alpha_plot}, as well as the quasi-linear scaling of such a maximum energy with $1/\kappa$.

To further corroborate this statement, we track where particles leave the jet, defining $z_{\rm f}$ as the value of $z$ where a particle trajectory crosses the surface of a cylinder that shares the jet's axis and has radius $15 \rj$ and length $80 \rj$.
The distribution of $z_{\rm f}$ is shown in Figure~\ref{zf}, which highlights how particles are more likely to escape in the transverse direction closer to the base of the jet for larger SGS, i.e., lower values of $\kappa$.
More precisely, the top panel shows the fraction of all the particles that leave the jet as a function of $z_f$, while the bottom panel only considers particles with sub-Hillas energies.
A comparison between the two panels reveals that for small values of $\kappa$ the bulk of sub-Hillas particles leaves the jet well before reaching the head;
without SGS, instead, most of the particles propagate to the jet's end (peak in the top panel).
Intermediate values of $\kappa$ bridge these two regimes.

It is also interesting to notice that the effect of SGS saturates before reaching the Bohm regime, as attested by the fact that curves for $\kappa=1$ and $\kappa=10$ are quite similar in Figures \ref{ratio}, \ref{alpha_plot}, and \ref{zf}.
For such a strong SGS, the sub-Hillas peaks converge to $\alpha \sim 0.7$ and particles tend to escape well before reaching the jet's head (top panel of Figure~\ref{zf}); 
we interpret this as the signature of an intrinsic limitation in maximum energy that applies to particles that escape at $z_{\rm f}<50 \rj$, i.e. the \emph{stochastic Hillas limit}, as explained below in \S\ref{subsec:stoch-hillas}.

\subsection{The importance of the stochastic Hillas limit}\label{subsec:stoch-hillas}
\citet{hillas84} discussed how stochastic (statistical) acceleration is characterized by an energy gain per cycle that competes with the particle escape probability, such that the escape time decreases as the particle energy increases. This, in turn, sets a stochastic Hillas limit, more easily achievable for particles that undergo stochastic acceleration, and generally smaller than the Hillas limit corresponding to the potential drop due to the motional electric field on the source diameter. As a result, the maximum energy of particles that undergo stochastic acceleration should depend on the scattering rate (i.e., on $\kappa$) and on the parameters ($V_r$, B, and transverse size) of both the spine and the cocoon.
Considering the balance between advection and diffusion outside of the spine of the jet expressed in Equation \ref{eq:diff}, we find that this stochastic Hillas limit sits at $\alpha_{\rm SHL} \sim 0.7/\kappa$ across the whole transverse jet region (spine+cocoon). Indeed, when we estimate such a limit using Equation \ref{eq:diff} for the relativistic spine's reference values ($\lambda = \rj$, $B \sim 8 B_0$, and $\langle V_r \rangle \sim 0.04$), we find that it is comparable with $\alpha_{\rm SHL}$.

From these considerations a simple picture arises: more SGS on one hand increases the fraction of accelerated particles, but on the other hand enhances the probability of leaving the jet sideways;
this leads to a pile-up at a maximum energy dictated by the stochastic Hillas limit $\alpha\lesssim 1$, which is intrinsically lower than the energy achievable by the particles that manage to make it throughout the full jet extent (Bottom Panel of Figure~\ref{zf}). 
In terms of acceleration mechanisms, we can conclude that \emph{espresso} acceleration (which happens even without SGS) is responsible for the energization of the highest-energy particles that a jet can produce, i.e., those that can probe the full potential drop.
Conversely, stochastic acceleration---which depends on the assumed level of SGS---can be relevant for the energization of lower-energy UHECR.
The more effective SGS is at accelerating particles, the more particles pile up close to the stochastic Hillas criterion and the flatter the overall UHECR spectrum.

Let us now study how acceleration occurs in different regions, namely the jet spine (high-$\Gamma$) and the shearing region at the jet-cocoon interface.

\subsection{Acceleration in high-$\Gamma$ regions}\label{highG}

\begin{figure*}
\centering
\includegraphics[width=0.48\textwidth,clip=false,trim= 0 0 0 0]{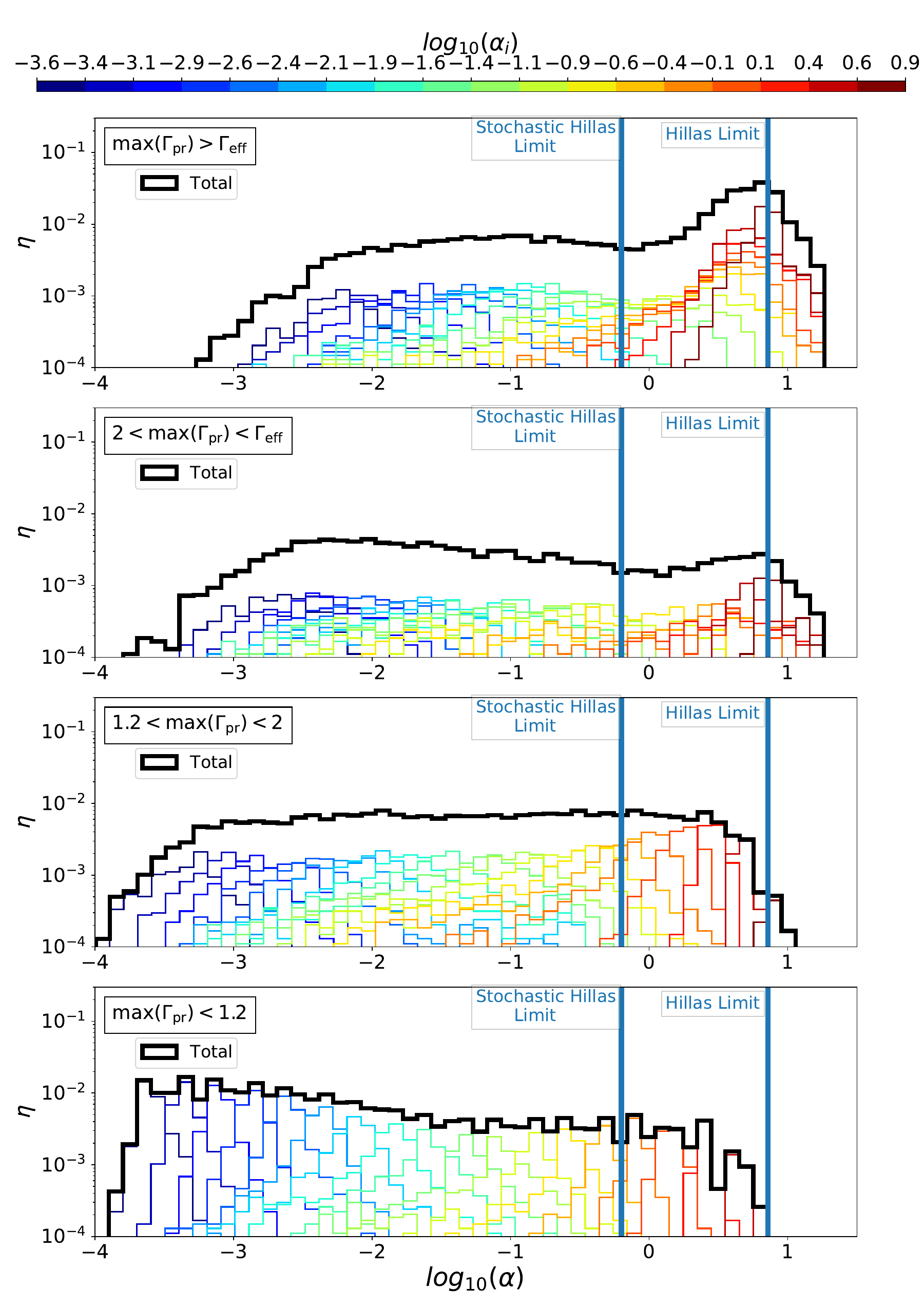}
\includegraphics[width=0.48\textwidth,clip=false,trim= 0 0 0 0]{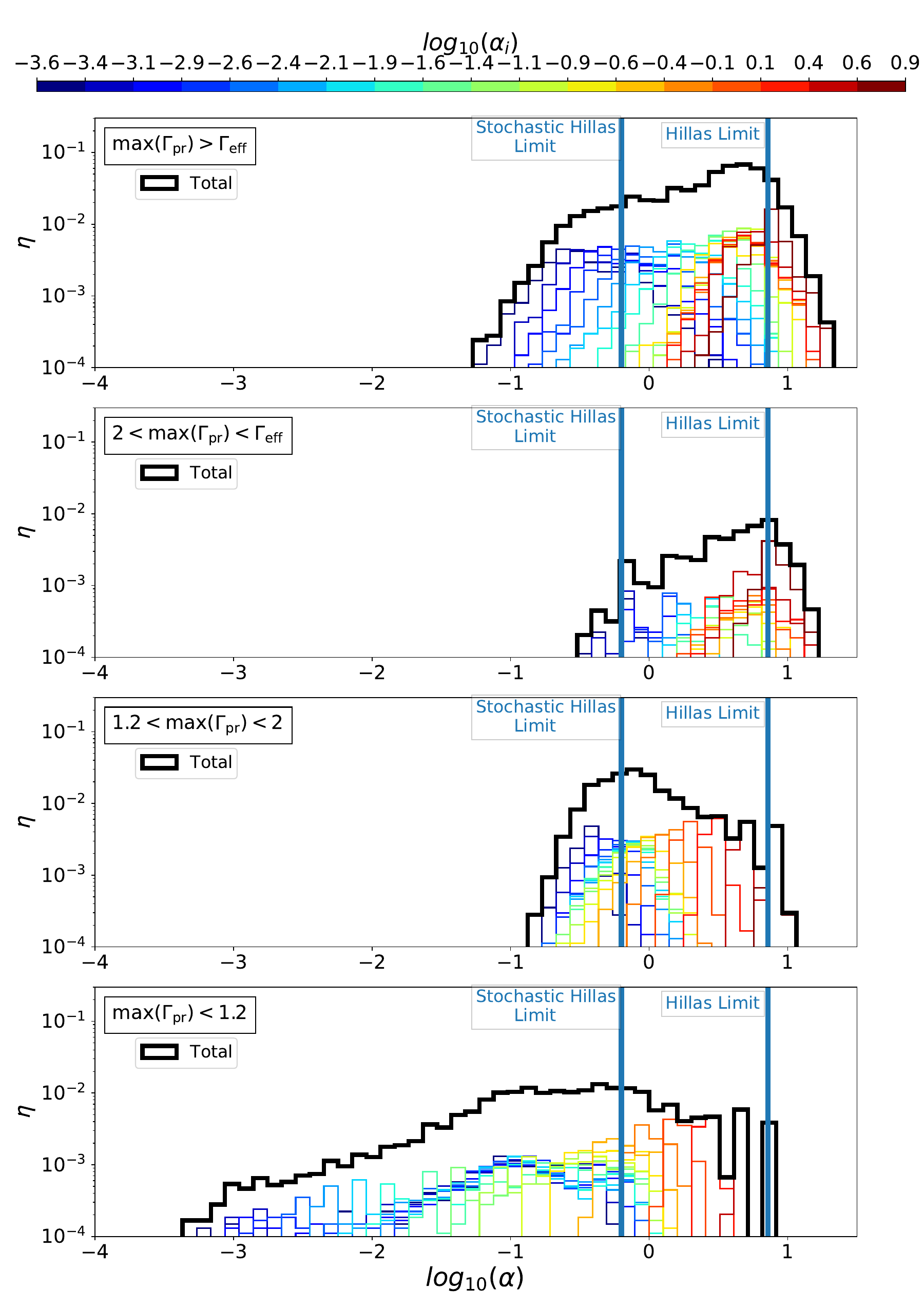}
\caption{Spectrum of accelerated particles separated according to the maximum Lorentz factor that they probe, $\maxg$ (top to bottom, as in the legends);
left and right panels correspond to $\kappa=\infty$ and $\kappa=1$, respectively.
Note that: 
1) the highest-energy CRs ($\alpha_{\rm f}\gtrsim 1$) systematically go through the most relativistic jet regions (top panels); 
2) adding SGS only incrementally enhances the flux of such particles (compare the top two panels);
3) adding SGS significantly boosts the energy of low-energy seeds (cold colors) that probe regions with $\Gamma_{\rm pr}\gtrsim 1.2$.
}
\label{Gamma_dependence}
\end{figure*}

Figure~\ref{Gamma_dependence} shows the impact of the maximum Lorentz factor, $\maxg$, that particles probe along their trajectories on the final spectrum, for the two extreme cases of $\kappa=\infty$ and $\kappa = 1$ (left and right panels, respectively);
the color code corresponds to the seed energy. 

Let us first focus on the highest-energy particles, those with final gyroradii $\alpha_{\rm f}\gtrsim 1$ (red histograms). 
Regardless of the amount of SGS, they typically probe the fastest jet regions with max$(\Gamma_{\rm pr})>\Gamma_{\rm eff}\sim 3.2$ (top panels in each column).
When maximally-effective SGS is added, the number of these particles remains substantially the same (right vs left panels).
This is again a manifestation that \emph{espresso} acceleration is sufficient to achieve the Hillas limit and that SGS may allow an incrementally larger fraction of the seeds to be reaccelerated via such a process. 
A similar conclusion can be drawn for particles that probe slower jet regions (top to bottom): adding SGS does not increase appreciably the amount of particles that achieve large final energies ($\alpha_{\rm f}\gtrsim 1$).

Finally, we can see that as $\maxg$ reaches 1.2 in the bottom panel, particles simply cannot reach the highest energies, even if with Bohm diffusion low-energy particles are still capable of gaining up to a factor of $\sim 50$ in energy, unlike in the $\kappa=\infty$ case. We can conclude here that particles are not likely to undergo \emph{espresso} in the lowest-$\Gamma$ regions where $\maxg <1.2$.

\subsection{Acceleration in low-$\Gamma$ regions}\label{lowG}

As mentioned in \S\ref{spec_sig}, in low-$\Gamma$ flows it is hard to unequivocally identify the most important acceleration mechanism, since multiple small \emph{espresso} shots are indistinguishable from stochastic acceleration. 
Therefore, we limit ourselves to analyzing trajectories and energy gains in comparison with \emph{espresso}'s ordered gyrations that we outlined in e.g., Figure~1 in MC19.
 Generally speaking, the most prominent effect of SGS in low-$\Gamma$ flows is to affect the mostly-ordered gyrations seen when $\kappa  = \infty$. 

\begin{figure}
\centering
\includegraphics[width=0.48\textwidth,clip=false,trim= 0 0 0 0]{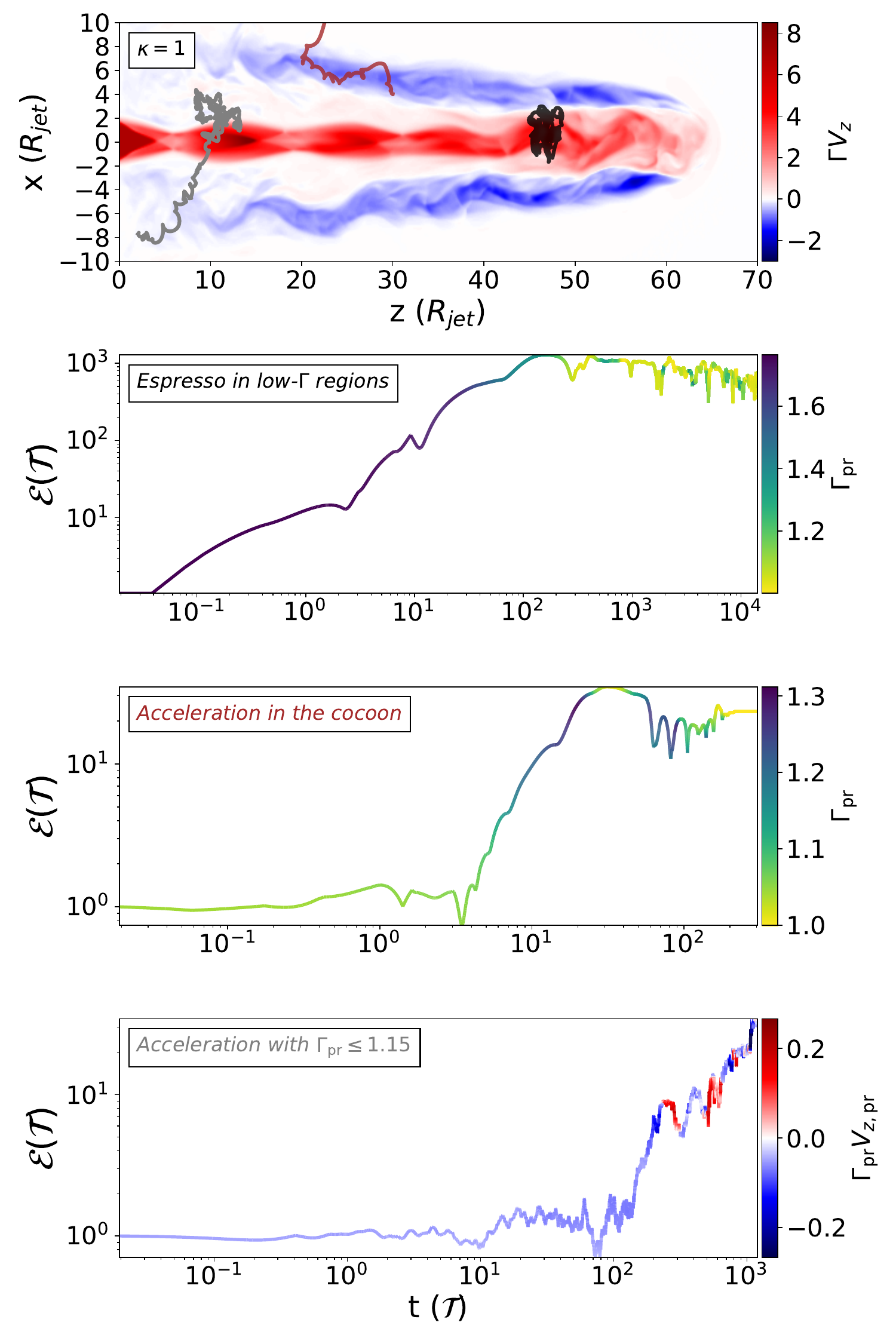}
\caption{2D projections of typical trajectories (top panel, plotted over the 4-velocity component $\Gamma v_z$ of the flow) and energy gains (bottom panels) for  particles that propagated only in low-$\Gamma$ regions, for $\kappa =1$. All particles escape the spine/cocoon system.
Second panel (black trajectory): energy gain as a function of $\mathcal T$, color coded with the instantaneous Lorentz factor probed, $\Gamma_{\rm pr}$. 
This is representative of particles that undergo multiple (3 in this case) \emph{espresso} shots in trans-relativistic regions around the jet spine.
Third panel (maroon): a particle accelerated in the jet backflow. 
Fourth panel (grey): a particle accelerated in sub-relativistic regions across the cocoon.}
\label{combined_traj}
\end{figure}

 Figure~\ref{combined_traj} shows three representative trajectories of particles that only probe low-$\Gamma$ regions before escaping, all for the case $\kappa=1$. 
The top panel shows their trajectories plotted over the 4-velocity component $\Gamma v_z$ of the jet, while the bottom panels show the energy gain as a function of time, with the color code corresponding to the instantaneous flow properties.

The black trajectory (second panel in Figure~\ref{combined_traj}) is representative of particles that probe the trans-relativistic region around the jet spine.
In just three gyrations, this particle, initialized with $\alpha_{\rm i} = 1.4 \times 10^{-3}$, experiences three very efficient \emph{espresso} shots---each corresponding to a maximum \emph{espresso} energy boost of a factor of $\sim 2\Gamma_{\rm pr}^2 \sim 8$ (Equation~11 in MC19)---with scattering events at roughly the peak of each gyration. 
Here the main role of SGS is to break the correlation between in- and out-going angles in the first cycles as, without the presence of SGS, this particle would not have gone through more than one ordered \emph{espresso} gyration. 
Finally, it lingers around in the cocoon ($\Gamma\sim 1$) before escaping with $\alpha_{\rm f} \sim 0.7 $, corresponding to the stochastic Hillas limit.

The maroon trajectory (third panel of Figure \ref{combined_traj}) is representative of particles that gain energy through multiple scattering events in the cocoon, without ever probing the relativistic spine.
Such a particle is initialized with $\alpha_{\rm i} = 0.025$ and gains up to a factor of $\sim 30$ in energy while probing $\Gamma_{\rm pr} \leq 1.3$. 
This particle escapes when it reaches the stochastic Hillas limit, with a total energy gain $\mathcal E\sim 30$.
We do not observe similar behavior in the absence of strong SGS, so it is fair to ascribe this kind of acceleration to stochastic acceleration in the jet backflow, as suggested, e.g., by \citet{osullivan+09,matthews+19}. 

Finally, the grey trajectory (bottom panel of Figure~\ref{combined_traj}) is representative of particles that only probe non-relativistic regions ($|v_z|\lesssim 0.2c$);
in this case, the trajectory is color-coded with the instantaneous 4-velocity component $\Gamma_{\rm pr} v_{\rm z, pr}$ of the flow. 
Such a particle, initialized with $\alpha_{\rm i} = 0.025$, experiences a mixture of energy gains and losses when crossing the shear layers as attested by the probed 4-velocity component, and finally exits the jet with an energy gain of $\sim 30$ without probing the relativistic spine. 
The last part of the trajectory can be ascribed to stochastic shear acceleration  \cite[e.g.][]{ostrowski98,ostrowski00,fang+18,kimura+18} due to the alternating sign of $v_z$.

Overall we find no evidence that particles that \emph{only} undergo stochastic acceleration fostered by the enhanced SGS can achieve boosts in excess of a factor of 50 in energy (bottom panel of Figure~\ref{Gamma_dependence}). 
Some particles, which start with small initial gyroradii can undergo larger boosts (as the black trajectory in Figure \ref{combined_traj}) if they probe relativistic regions;
all of these particles, though, make it only to the stochastic Hillas limit.

Our considerations are drawn by examining a large but finite number of particles, which means that we cannot exclude the existence of trajectories along which particles may be accelerated up to the longitudinal limit without undergoing any \emph{espresso} cycle;
yet, we can quantitatively assess that this is not common.

We may summarize these findings by saying that, while adding SGS allows particles to be stochastically accelerated also in the cocoon or at the jet interface (where there is free energy in the form of shear, wake shocks, and turbulence), most of the acceleration is bound to occur in the relativistic spine. The particles that are accelerated to the highest energies are those that manage to probe most of its extent, and a more efficient scattering may  hinder the process by enhancing lateral escape.

\subsection{A more elaborate diffusion prescription}

\begin{figure}
\centering
\includegraphics[width=0.45\textwidth,clip=false, trim= 10 0 0 0]{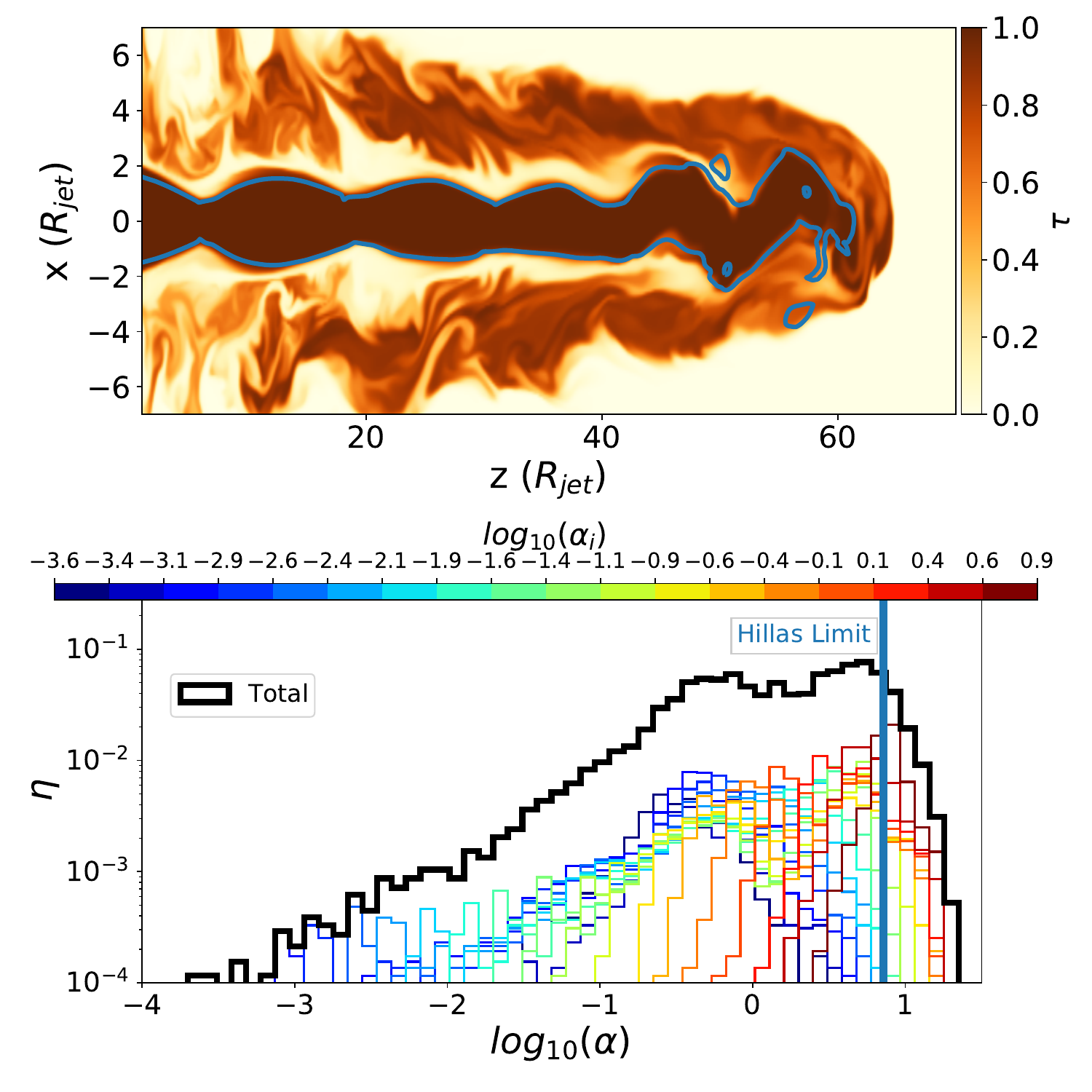}
\caption{Top Panel: 2D cut at $y=0$ of $\tau$, a tracer of the relative abundance of jet/ambient material. 
The blue contour plot delimits the spine of the jet, defined as the region with $\Gamma > 2$. 
Bottom Panel: As in Figure \ref{alpha_plot}, but with Bohm diffusion in the spine and galactic-like diffusion in the cocoon, à la \citet{kimura+18}. 
Note the similarity with the bottom panel of Figure~\ref{alpha_plot} ($\kappa = 1$).}
\label{kimura}
\end{figure}

The calculations above showed how UHECR spectra change when simple prescriptions for SGS are introduced. 
In reality, pitch-angle scattering depends on the level of turbulence at scales resonant with each particle, and quite different magnetic fluctuations are expected in the jet, the cocoon, and the ambient medium.

For instance, \citet{kimura+18} recently studied UHECR stochastic acceleration  in idealized jets via Monte Carlo simulations with different scattering rates in the cocoon and in the jet.
In particular, they assumed Bohm diffusion in the jet (corresponding to our spine) only, while outside the mean free path was chosen as $\lambda_i = l_{\rm c} (\mathcal{R}/\mathcal{R}_{\rm c})^\delta$ where $l_{\rm c}$ is the coherence length of the local magnetic turbulence and $\mathcal{R}_c$ the gyroradius for which $\mathcal{R}(E,q)=l_{\rm c}$.
For $\mathcal{R} < \mathcal{R}_{\rm c}$, Kolmogorov turbulence is assumed ($\delta = 1/3$); otherwise, particles should only see turbulence on scales much smaller than their gyroradii, which leads to $\delta = 2$. The coherence length $l_{\rm c}$ is set to be 3\% of the radius of the cocoon.

We now investigate a similar prescription for particles propagated in our benchmark simulation and show the results in Figure \ref{kimura}.
More precisely, we define the spine of the jet as the region with $\Gamma \geq 2$ (delimited by the blue contour plot in the top panel) and the cocoon as the region encompassed by the shocked jet material, traced by the value of $\tau$, a scalar quantity that describes the mixing between jet ($\tau=1$) and ambient ($\tau=0$) material.

The bottom panel of Figure~\ref{kimura} shows that the spectrum of reaccelerated particles is very similar to the case with Bohm diffusion ($\kappa = 1$) in Figure~\ref{alpha_plot}. 
In particular, the peak at $\alpha_{\rm f}\lesssim 1$, corresponding to the stochastic Hillas limit, is quite prominent.

This  suggests that CR reacceleration in AGN jets is controlled by the effective scattering rate in the highly-relativistic regions and weakly dependent on the details of how particles diffuse in the cocoon. 
We stress again that the actual amount of SGS that is realized in realistic environments is highly uncertain, but it is conceivable that in the jet SGS may be enhanced with respect to the ambient medium, which should be described by the galactic diffusion coefficient.

\section{espresso in realistic environments}\label{realistic}

Let us discuss now how our findings apply to different AGNs from which UHECRs may originate; 
we limit ourselves to radio-loud AGNs, which are the only ones energetic enough to potentially supply the UHECR luminosity  \citep[e.g.,][and references therein]{fang+18,kimura+18,mbarek+19}. 
Within radio-loud candidates, we distinguish between FR-I and FR-II jets \citep{fanaroff+74}, whose properties are likely determined by both the engine luminosity and the density profile of the ambient medium \citep{tchekhovskoy+16}.
FR-I jets are typically decelerated to trans-relativistic velocities within 1 kpc \citep[e.g.,][]{wardle+97,arshakian+04,mullin+09}, while FR-II jets are more powerful and can sustain $\Gamma>10$ flows over tens of kpc and extend up to hundreds of kpc before they are dissipated \citep[e.g.,][]{sambruna+02,siemiginowska+02,tavecchio+04,harris+06}.
Strictly speaking, our benchmark simulation should be more similar to a FR-I radio galaxy since the ambient density is homogeneous, rather than rapidly decreasing as in a FR-II case, and the jet is decelerated on a scale of a few tens of $\rj$. A newer family of radio galaxies dubbed FR-0s has also emerged to represent the bulk of the radio-loud AGN population in the near-universe with redshifts $\lesssim 0.05$ \citep[e.g.][]{baldi+18a,baldi+18b,torresi+18,garofalo+19}. With an extent that can reach 3kpc, the only notable difference between FR-0s and extended FR-Is is the former's lack of extended radio emission \citep{baldi+18b,garofalo+19}. FR-0s could be associated with early-type galaxies that can evolve into FR-Is provided that their central black hole's spin is boosted with increasing accreting matter, but FR-0s do not need to be young objects and could also be associated with decelerated FR-IIs with decreasing power considering their important extent \citep{baldi+18a,baldi+18b,garofalo+19}. \citet{merten+21} suggested that FR-0 jets can accelerate particles to UHECR levels through stochastic shear acceleration. While this possibility cannot be ruled out, the fact that for low bulk Lorentz factors the stochastic Hillas limit is significantly more stringent than the one due to the motional electric field  suggests that UHECR sources may be limited to FR-I and FR-II jets.

As discussed in \S\ref{init_traj}, the highest energies are achieved by particles that can penetrate into the jet spine;
therefore, we expect the extent of the jet to control the highest achievable energy (see Equation~16 in MC19 and \S\ref{subsec:stoch-hillas}).
This favors FR-II jets, which are both faster and longer, as the candidate sources of UHECRs up to $10^{20}$eV;
this conclusion was already drawn by MC19, but here it is reinforced because we have shown that the highest achievable energies are independent of the level of SGS (see Figure~\ref{alpha_plot}).

SGS is arguably more important in FR-I jets with relatively small Lorentz factors, where one/two-shot \emph{espresso} acceleration is not enough to produce the highest-energy CRs starting from galactic-like CRs. 
As shown in section~\ref{lowG}, with quasi-Bohm diffusion, particles can go through multiple acceleration events in low-$\Gamma$ regions and gain up to a factor of $10^4$ in energy, thereby reaching the stochastic Hillas limit (see \S\ref{subsec:stoch-hillas}).

This suggests that FR-I galaxies may also contribute to the bulk of UHECRs around $10^{18}-10^{19}$eV, with a spectrum which may be more or less flat depending on the amount of SGS (see \S\ref{sec_alpha}).

Of course this includes nearby AGNs, such as Centaurus A and M87, as potential sources of UHECRs.
Note that they may not look like hotspots in the map of UHECR directions of arrival not only because of particle deflections in the intergalactic medium, but also because their contribution may be swamped in the flux from all the other AGNs on cosmological scales.
Such an effect (à la Olbers' paradox) may arise because UHECR protons with energies $\gtrsim 10^{18}$eV may travel almost unhindered across the whole universe;
therefore, in this energy window we should expect also the contribution from the most powerful blazars and flat spectrum radio quasars at the cosmological peak of AGN activity, at redshift $z\sim 1-2$. 
To explain the highest CR energies one would still need local, i.e., within $\sim 200$ Mpc ($z\lesssim 0.05$), highly-relativistic sources; 
the catalogue of such AGNs is very likely incomplete, but at least a few sources are present \citep[][]{caprioli18}.

Finally, we recognize that the spectra of heavy ions may be affected by photodisintegration in and around the UHECR sources as they escape from the jet. 
A thorough analysis of the role of UHECR losses within the \emph{espresso} framework, along with the expected flux of neutrinos produced in AGNs, will be presented in a forthcoming paper.

\section{Conclusions}
In MC19 we tested the \emph{espresso} framework \citep{caprioli15} as a mechanism for the acceleration of UHECRs in AGN jets.
In particular, we used a bottom-up approach in which test particles are propagated in a 3D MHD simulation, finding that the model is consistent with the current UHECR phenomenology in terms of spectra, chemical composition, and anisotropy, independently of the scattering rate.

In this paper, we further such an investigation by including subgrid scattering (SGS) to characterize the role of the small-scale magnetic irregularities that are not captured in MHD simulations. 
Our framework is mechanism-agnostic, i.e., seed trajectories are integrated via standard particle-in-cell techniques \citep[e.g.,][]{birdsall+91}, augmented with a Monte Carlo treatment of pitch-angle scattering.
We bracket our ignorance of the actual level of scattering by spanning from cases with no SGS (à la MC19) to Bohm diffusion;
this allows us to assess the relative importance of \emph{espresso} and stochastic acceleration in accelerating UHECRs in AGN jets. 
Our results can be summarized as:

\begin{enumerate}
    \item Adding SGS fosters stochastic acceleration, which can contribute to initially energize the seeds, enabling some of them to penetrate in the higher $\Gamma$ regions and get \emph{espresso} accelerated (Figure~\ref{traj}). 
    
    \item SGS helps to break the correlation between the in-going and out-going angles in \emph{espresso} cycles, thereby increasing the energy gain in each gyration in the jet spine.
    This leads to larger average energy gains (Figure \ref{max_gain}): two or three \emph{espresso} shots are common (i.e., experienced by $\gtrsim 10\%$ of the seeds) for large levels of SGS.
    The maximum energy gain observed for Bohm diffusion is a factor of $10^2$ larger than without SGS (Figure \ref{ratio}).
    
    \item However, the importance of SGS is limited to low-energy UHECRs. The highest-energy particles are invariably \emph{espresso}-accelerated and the spectrum of the particles that reach the Hillas limit is independent of the level of SGS (see Figure \ref{Gamma_dependence}).
    
    \item Figure \ref{Gamma_dependence} also shows that the highest-energy particles always probe the most relativistic jet regions. 
    Particles that only experience stochastic acceleration in the cocoon or in non-relativistic regions experience a boost of a factor of $\sim 50$ at most and do not contribute significantly to the UHECR flux.
    
    \item Introducing SGS increases the probability of accelerated particles to escape the jet sideways (Figure \ref{zf}).   
    This leads to a pile-up close to the stochastic Hillas limit (Figure \ref{alpha_plot}), achieved at the energy for which the diffusion length becomes comparable with the jet transverse size (Equation \ref{eq:diff}).
    For typical jet aspect ratios, such a limit is lower than the Hillas limit associated with the motional electric field (See \S\ref{subsec:stoch-hillas}) and is dependent on the scattering rate, consistent with the fact that adding SGS does not change the UHECR spectrum at the highest energies.  
    
    \item The combination of the effects above is such that adding more and more SGS produces flatter and flatter UHECR spectra, which explains the results obtained assuming Bohm diffusion, e.g, by \citet{kimura+18}; 
    therefore, small-scale diffusion may be responsible for producing the rather flat UHECR spectra that are preferred by propagation models \cite[e.g.,][]{aloisio+14,gaisser+13,taylor14}. 
     
    \item In terms of AGNs as potential UHECR sources, SGS fosters the reacceleration of a large fraction of seed CRs, but does not change the maximum achievable energy, which is expected to be larger for more extended jets. 
    This suggests that typical radio-bright AGNs should be able to accelerate particles at least to the stochastic Hillas limit, potentially filling the transition between Galactic and extra-galactic CRs and contributing to the lowest-energy UHECRs.
    
    \item \emph{Espresso} acceleration in powerful and extended FR-II jets, where the longitudinal Hillas criterion is maximized (See Equation 16 in MC19), remains the lead candidate for the production of the highest-energy CRs, independently of our poor knowledge of the actual CR diffusion rate in AGN jets.
\end{enumerate}

\acknowledgements
We kindly thank the referee for their thorough comments. Simulations were performed on computational resources provided by the University of Chicago Research Computing Center.
This research was partially supported by NSF through grants PHY-1748958 and PHY-2010240.



\bibliography{Total}
\bibliographystyle{aasjournal.bst}

\end{document}